\documentclass{webofc}
\usepackage[varg]{txfonts}   % Web of Conferences font
\newcommand{\be}{\begin{equation}}
\newcommand{\ee}{\end{equation}}

\makeatletter
\newcommand*{\rom}[1]{\expandafter\@slowromancap\romannumeral #1@}
\makeatother
\makeatletter

\newcommand{\Rmnum}[1]{\expandafter\@slowromancap\romannumeral #1@}
\woctitle{13th International Symposium on Origin of Matter and Evolution of Galaxies}
\begin{document}
\title{A Chemical Evolution Model for the Fornax Dwarf Spheroidal Galaxy}
\author{Zhen Yuan\inst{1,2}\fnsep\thanks{\email{yuan@physics.umn.edu}} \and
        Yong-Zhong Qian\inst{1,3} \and
        Yi Peng Jing\inst{2}
        % etc.
}

\institute{School of Physics and Astronomy, University of Minnesota, Minneapolis, 
MN 55455, USA
\and
Center for Astronomy and Astrophysics, Department of Physics and Astronomy, 
Shanghai Jiao Tong University, Shanghai 200240, China
\and
Center for Nuclear Astrophysics, INPAC, Department of Physics and Astronomy, 
Shanghai Jiao Tong University, Shanghai 200240, China
}

\abstract{%
Fornax is the brightest Milky Way (MW) dwarf spheroidal galaxy 
and its star formation history (SFH) has been derived from observations. 
We estimate the time evolution of its gas mass and net inflow and outflow 
rates from the SFH using a simple star formation law that relates
the star formation rate to the gas mass. We present a chemical evolution model
on a 2D mass grid with supernovae (SNe) as sources of metal enrichment. We 
find that a key parameter controlling the enrichment is the mass $M_x$ of the
gas to mix with the ejecta from each SN. The choice of $M_x$ depends on the 
evolution of SN remnants and on the global gas dynamics. It differs between 
the two types of SNe involved and between the periods before and after 
Fornax became an MW satellite at time $t=t_{\rm sat}$. Our results indicate
that due to the global gas outflow at $t>t_{\rm sat}$, part of the ejecta from 
each SN may directly escape from Fornax. Sample results from our model 
are presented and compared with data.
}
\maketitle

\section{Introduction}
\label{intro}
With the advent of large telescopes and supercomputers, studies of 
galaxy formation and evolution have entered a new era. In the meantime,
there has also been major progress in our understanding of stellar
explosion and nucleosysnthesis. Elemental abundances in stars form an 
important link between galactic and stellar processes. Because the surface
abundances of long-lived low-mass stars record the composition of their
birth environments, they serve as fossils of galactic chemical evolution. 
By studying abundances in stars of different ages in an individual galaxy,
we can obtain valuable information on its formation and evolution. 
Here we focus on nearby dwarf spheroidal galaxies (dSphs), which are 
considered to be surviving building blocks of larger galaxies in the 
framework of hierarchical structure formation. Understanding how dSphs
evolve is crucial to unraveling the formation and evolution of larger 
galaxies such as the Milky Way (MW). Unlike larger galaxies that have
undergone frequent mergers, dSphs have a smoother evolution,
which makes them a good laboratory to study various physical processes
on galactic scales.

As an example of the general approach, we first estimate the time evolution
of the gas mass and net inflow and outflow rates from the star formation 
history (SFH) for Fornax, the brightest MW dSph. We then build a model for 
its chemical evolution and compare the results with stellar abundance data. 
Observations strongly indicate that chemical enrichment in Fornax remains
inhomogeneous until the very end of its SFH despite its relatively small size. 
Using nucleosynthetic yields from both core-collapse (CCSNe) and Type Ia 
supernovae (SNe Ia), we numerically simulate the stochastic and 
inhomogeneous mixing of newly-synthesized elements in a 2D disk system. 
By comparing the resulting abundance distributions with data on Fornax, we 
find that the mixing depends on large-scale gas flows and the environments 
surrounding CCSNe and SNe Ia.

\section{Global Gas Dynamics of Fornax}
We assume a star formation law of the Schmidt-Kennicutt form:
\be
\frac{\Sigma_{\rm SFR}}{M_\odot\ {\rm yr}^{-1}\ {\rm kpc}^{-2}}=
C\left(\frac{\Sigma_g}{10\,M_\odot\ {\rm pc}^{-2}}\right)^\alpha,
\ee
where $\Sigma_{\rm SFR}$ is the star formation rate (SFR) per unit area,
$\Sigma_g$ is the surface density of gas, and $C$ and $\alpha$ are
constants. We consider that star formation in Fornax occurred in a uniform 
disk, the area $A_{\rm disc}(t)$ of which grew until Fornax became an MW 
satellite at time $t=t_{\rm sat}$ and then stayed fixed until the end of star 
formation. Under the above assumptions, the global SFR $\psi(t)$ can be 
written as
\be
\psi(t)=\lambda_*(t)\left[\frac{M_g(t)}{M_\odot}\right]^\alpha,
\label{eq-sfr}
\ee
where $M_g$ is the total mass of gas in the disk, and
\be
\lambda_*(t)=C\left[\frac{10\ {\rm kpc}^2}{A_{\rm disc}(t)}\right]^{\alpha-1}
\times10^{1-8\alpha}\ M_\odot\ {\rm yr}^{-1}.
\ee
As $A_{\rm disc}(t)=A_{\rm disc}(t_{\rm sat})$ for $t\geq t_{\rm sat}$, the above
equation can be rewritten as
\be
\lambda_*(t)=\left\{\begin{array}{ll}
\lambda_*(t_{\rm sat})[A_{\rm disc}(t_{\rm sat})/A_{\rm disc}(t)]^{\alpha-1},
&t<t_{\rm sat},\\
\\
C[10\ {\rm kpc}^2/A_{\rm disc}(t_{\rm sat})]^{\alpha-1}
\times10^{1-8\alpha}\ M_\odot\ {\rm yr}^{-1},&t\geq t_{\rm sat}.
\end{array}\right.
\label{eq-lambda}
\ee
We estimate $A_{\rm disc}(t_{\rm sat})\sim 10$~kpc$^2$ from the radial
distribution of stars in the present-day Fornax \cite{deboer-f}. In accord with
values of $C$ and $\alpha$ derived from observations of star formation
in nearby faint dwarf irregular galaxies \cite{roychow}, we take $\alpha=1.5$ 
and  $\lambda_*(t_{\rm sat})=8\times10^{-15}\,M_\odot$~yr$^{-1}$ for the 
presentation below.

\begin{figure}[ht!]
\includegraphics[width=0.5\textwidth]{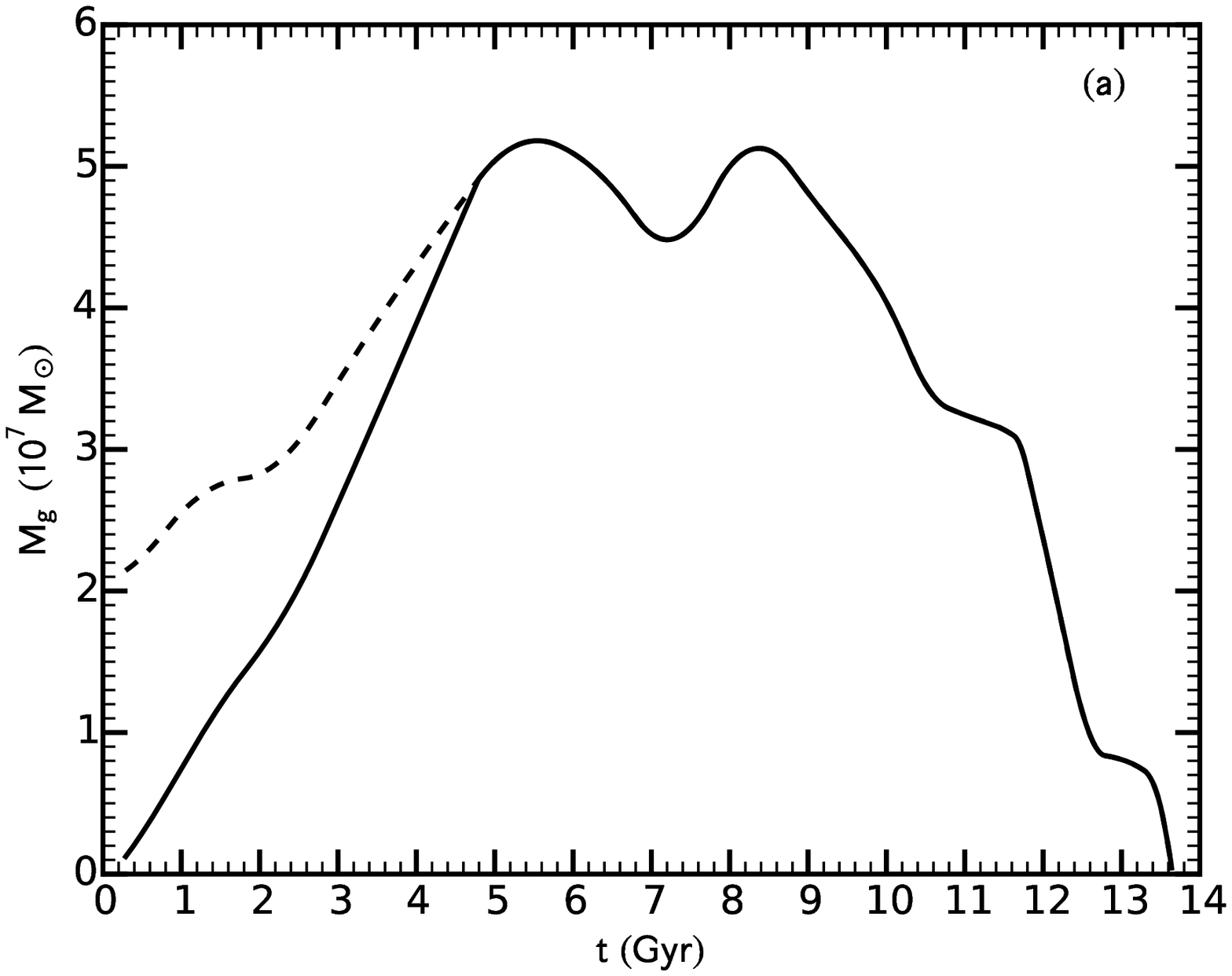}%
\includegraphics[width=0.5\textwidth]{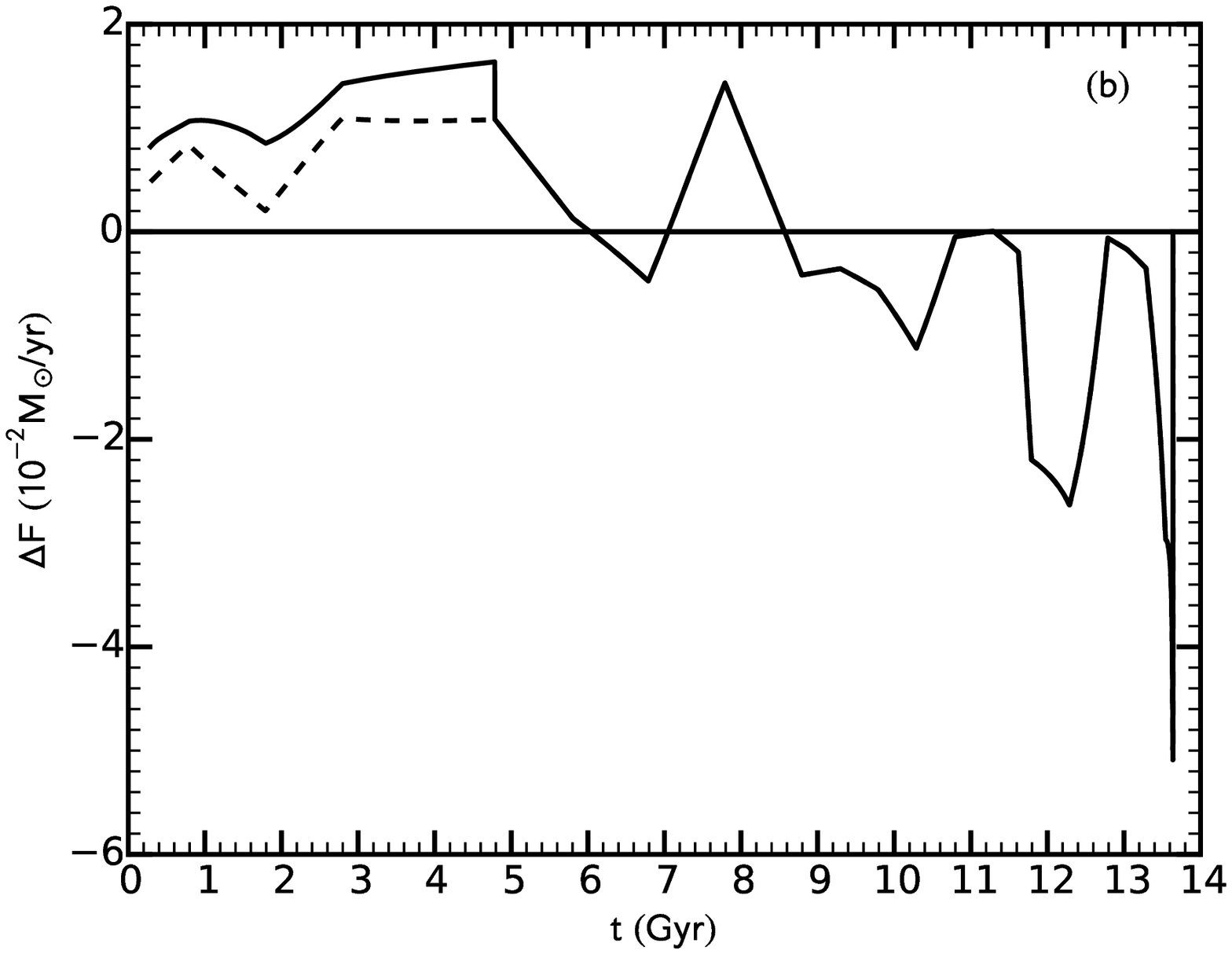}
\caption{Evolution of (a) the gas mass $M_g$ and (b) the net gas flow 
rate $\Delta F$ as functions of time $t$ for Fornax. We estimate that 
Fornax became an MW satellite at $t_{\rm sat}\approx 4.8$~Gyr. The 
solid curves correspond to a star-forming disk that grew until $t_{\rm sat}$ 
and then stayed fixed. The dashed curves assume the same disk size 
before and after $t_{\rm sat}$.}
\label{fig-fg-disk}
\end{figure}

We use Eq.~(\ref{eq-sfr}) to estimate $M_g$ from Fornax's SFH as
\be
M_g(t)=\left[\frac{\psi(t)}{\lambda_*(t)}\right]^{1/\alpha}M_\odot.
\ee
We further estimate the net gas flow rate as 
\be
\Delta F(t)\equiv\frac{dM_g}{dt}+\psi(t).
\label{eq-df}
\ee
We expect a net inflow, i.e., $\Delta F>0$ for $t<t_{\rm sat}$ as gas was 
accreted along with dark matter during the growth of the halo hosting Fornax. 
The halo stopped growing when it became an MW satellite. Subsequently, its 
tidal interaction with the MW and the ram pressure associated with its orbital
motion worked together to remove gas from Fornax. So we expect a net
outflow, i.e., $\Delta F<0$ for most of the time $t>t_{\rm sat}$. To estimate
$t_{\rm sat}$, we first assume a fixed $\lambda_*(t)=\lambda_*(t_{\rm sat})$
for all $t$. Using the data on $\psi(t)$ \cite{deboer-f}, we find that the growth 
of $M_g$ slowed down greatly (Fig.~\ref{fig-fg-disk}a) and $\Delta F$ started 
to decrease towards zero for the first time (Fig.~\ref{fig-fg-disk}b) at 
$t\approx 4.8$~Gyr. If we take this time as the estimate for $t_{\rm sat}$,
then $d\psi/dt>0$ and $d\lambda_*/dt<0$ ensure that $\Delta F>0$ for 
$t<t_{\rm sat}$ even when we allow $\lambda_*(t)$ to change as in 
Eq.~(\ref{eq-lambda}). We adopt $t_{\rm sat}\approx 4.8$~Gyr.

Using the median halo growth history given by Ref.~\cite{zhao}, we 
estimate that the Fornax halo grew to a total mass of 
$M_h\approx 1.8\times 10^9\,M_\odot$ at $t_{\rm sat}\approx 4.8$~Gyr 
in order to obtain the mass enclosed within
the half-light radius inferred from observations \cite{wolf}. Taking 
$A_{\rm disc}(t)/A_{\rm disc}(t_{\rm sat})=
[r_{\rm vir}(t)/r_{\rm vir}(t_{\rm sat})]^2$ for $t<t_{\rm sat}$, where 
$r_{\rm vir}$ is the virial radius of the Fornax halo, we show the 
corresponding evolution of $M_g$ and $\Delta F$ in Fig.~\ref{fig-fg-disk}
(see Refs.~\cite{yuan,yuan-th} for more details).
The full evolution of $M_g$ and $\Delta F$ will be used to build a 
chemical evolution model for Fornax below.

\section{Simulating Chemical Evolution of Fornax}
Fornax is $\sim 10^3$ times smaller than the MW and has a prolonged 
SFH that lasted almost the age of the universe. Fornax is also chemically
much more inhomogeneous \cite{deboer-f} than the MW disk although
both systems have evolved for approximately the same amount of time. 
This apparently counterintuitive result can be understood from the 
efficiency for mixing of the interstellar medium by SNe in a gaseous
system of mass $M_g$. Assuming that a total mass $M_x$ of the 
interstellar medium is mixed by an SN, we estimate that the rate for
the fraction of gas mixed is $f_{\rm mix}\propto\psi M_x/M_g\propto 
M_g^{\alpha -1}$, which increases with $M_g$ because $\alpha>1$.
This suggests that the larger a system is, the more quickly it would 
become chemically homogeneous. In addition, the rate for enrichment 
of an element E can be estimated as $f_{\rm E}\propto\psi
\langle Y_{\rm E}\rangle/M_g\propto M_g^{\alpha -1}$, where 
$\langle Y_{\rm E}\rangle$ is the average SN yield of E.
Consequently, a larger system such as the MW is more chemically
enriched than Fornax and other dSphs \cite{tol09} even if the
duration of enrichment is the same. 

To simulate the chemical evolution of Fornax, we use a 2D mass grid
with $n=10^3\times 10^3$ boxes to approximate its gaseous disk for
star formation. The number of boxes filled with gas changes with time
according to the gas mass $M_g$ shown as the solid curve in 
Fig.~\ref{fig-fg-disk}a. Each gas box contains a fixed mass 
$m_0=M_{g,{\rm max}}/n$, where $M_{g,{\rm max}}$ is the maximum
gas mass. We take time steps of 0.2~Gyr each.
For the $i$th time step $t_{i-1}\leq t<t_i$ ($i\geq 1$), we first 
treat the gas inflow or outflow by adding or dropping gas boxes, 
respectively, so that the number of gas boxes is $n_i=M_{g}(t_i)/m_0$. 
We assume that the inflow is metal free during the pre-satellite phase 
($t<t_{\rm sat}$) and for $t_{\rm sat}< t\lesssim 6$~Gyr during the 
satellite phase. The inflow for $7\lesssim t\lesssim 8.6$~Gyr consists of 
reaccreted gas and its metallicity is calculated from our model.
We calculate the numbers $n_{{\rm CC},i}$ and $n_{{\rm Ia},i}$ of 
CCSNe and SNe Ia, respectively, 
for the $i$th time step based on the SFH. 
The fraction of stars resulting in CCSNe is estimated 
from the initial mass function given by Ref.~\cite{kroupa}. The birth rate 
for progenitor systems of SNe Ia is specified as a fraction of the CCSN 
rate. The delay between the birth of an SN Ia progenitor system and its 
explosion is assumed to follow the distribution given by Ref.~\cite{maoz}.
We then randomly pick $n_{{\rm CC},i}$ and $n_{{\rm Ia},i}$ from a total 
of $n_i$ boxes as sites of CCSNe and SNe Ia, respectively. Both types 
of SNe inject newly-synthesized elements into the surrounding gas boxes. 
The mixing and chemical enrichment of the gas surrounding each SN is 
described below. To finish the $i$th time step, we randomly sample 
$n_{s,i}$ boxes, where $n_{s,i}$ is the number of stars that were formed 
during this time step and survive to the present based on the SFH and
stellar lifetimes. The above procedure is repeated for the entire SFH. 

The elemental yields of CCSNe and SNe Ia are taken from 
Refs.~\cite{heger} and \cite{maeda}, respectively. A specific SN enriches
and mixes the gas in a total of $n_x=M_x/m_0$ surrounding boxes. 
The resulting mass fraction of element E in these boxes is
\be
X'_{{\rm E},j}=\frac{m_0\sum_{k=1}^{n_x} X_{{\rm E},k}+Y_{\rm E}}{M_x},
\ j=1, 2, \cdots, n_x,
\ee
where $X_{{\rm E},k}$ is the mass fraction of E in the $k$th box
before the SN occurs and $Y_{\rm E}$ is the yield of the SN.
In view of the dominance of the outflow during the satellite phase, 
$Y_{\rm E}$ is reduced for this phase by a factor of 2 or 4 from the
theoretical yield for a CCSN or SN Ia, respectively, to account for 
approximately the direct loss of metals from each SN. 
The total mass $M_{x}$ of the gas to mix with
each SN is estimated from the results in Ref.~\cite{thornton}. As 
CCSNe are associated with active star-forming regions of high 
densities but SNe Ia tend to occur in regions of low densities, we 
set the ratio of the mixing masses for each SN Ia and CCSN to be 
$M_{x,{\rm Ia}}/M_{x,{\rm CC}}=5$ throughout the SFH. The
mixing mass is also affected by the global gas flow. During the
phase dominated by the inflow, cold dense gas falls into the 
system while hot ``bubbles'' are blown out by SNe. This would 
drive significant turbulence on the galactic scale \cite{kgf15}
and result in widespread mixing of gas. In contrast, SN bubbles 
cannot survive for long during the phase dominated by the outflow,
which suggests inefficient mixing for this phase. Based on the above 
discussion, we set $M_{x,{\rm Ia}}=10^5\,M_{\odot}$, 
$M_{x,{\rm CC}}=2\times10^4\,M_{\odot}$ during the pre-satellite phase, 
and $M_{x,{\rm Ia}}=5\times10^3\,M_{\odot}$, 
$M_{x,{\rm CC}}=10^3\,M_{\odot}$ during the satellite phase.

\section{Results}
A snapshot of the chemical composition in each gas box is taken at the 
end of each time step. These snapshots describe the inhomogeneous 
chemical evolution of Fornax. In particular, snapshots of gas boxes 
sampled by simulated stars that were formed during each time step 
and survive until today can be compared with observations.
Here we only present sample results on [$\alpha$/Fe] vs. [Fe/H],
where $\alpha$ stands for Mg, Si, Ca, and Ti (see Ref.~\cite{yuan-th} 
for more details). 

\begin{figure*}[ht!]
\hskip-0.1cm
\includegraphics[width=0.98\textwidth]{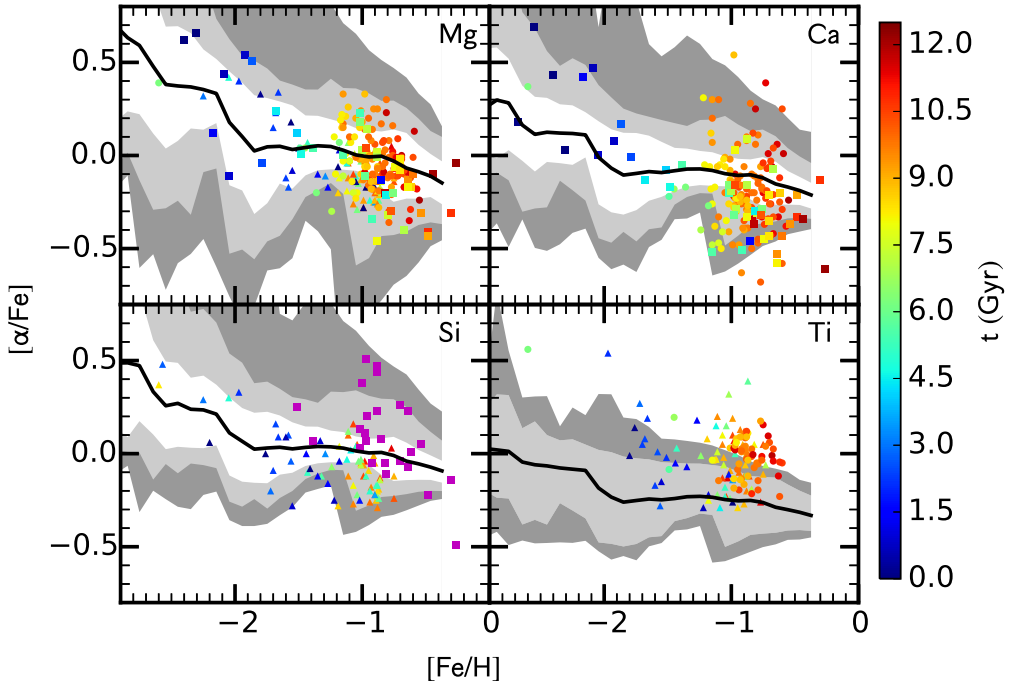}
\caption{Comparison of simulation results with stellar data on [$\alpha$/Fe]
vs. [Fe/H] for Fornax. The solid curves represent the average [$\alpha$/Fe] 
values as functions of [Fe/H] from simulations. The central white,
the light grey, and the dark grey regions indicate (cumulatively) the
$1\sigma$, $2\sigma$, and $3\sigma$ confidence regions, respectively,
for Mg, Si, and Ca. Only the $2\sigma$ and $3\sigma$ confidence 
regions are shown for Ti. 
Data are taken from Refs.~\cite{deboer-f,letarte,kirby} (circles),
\cite{lemasle} (squares), and \cite{hendricks} (triangles). Colors 
indicate the formation time of stars except for the purple squares, for which
no such information is available. Note the general agreement for Mg, Si, 
and Ca as well as the offset for Ti between simulation results and data.
The offset for Ti may be due to systematic underestimates of Ti yields
by CCSN models for some unknown reasons or to systematic 
uncertainties in measurements of Ti abundances.}
\label{fig-alpha}
\end{figure*}

We compile the simulated stars in the increasing order of their [Fe/H] 
values. At a specific [Fe/H] value, we calculate the algebraic averages of
[$\alpha$/Fe] and show these as the solid curves in Fig.~\ref{fig-alpha}. 
We also determine the ranges of [$\alpha$/Fe] 
covered by 68\%, 95\%, and 99.7\% of the stars, respectively.
These ranges correspond to the $1\sigma$, $2\sigma$ and $3\sigma$ 
confidence regions, and are indicated (cumulatively) by the central white, 
the light grey, and the dark grey regions, respectively, in Fig.~\ref{fig-alpha} 
for Mg, Si, and Ca. The stellar data \cite{deboer-f,letarte,kirby,lemasle,hendricks} 
are also shown for comparison. As most of the data on Mg, Si, and Ca lie 
within the 95\% confidence regions, the scatter of the data can be 
approximately accounted for by our model. 
In addition, the general trends of these data 
can be described by the solid curves. We note that the agreement between 
our results and the data is the best for Mg, which also happens to be the 
best measured $\alpha$ element. There are some data on Si and Ca that 
lie in the dark grey regions. How representative these data are should
be determined by high-quality observations of more stars.

The solid curve for Ti lies below most of the data. In addition, because Ti 
and Fe are coproduced by explosive nucleosynthesis with similar yield ratios 
for all CCSNe, the $1\sigma$ confidence region for [Ti/Fe] vs. [Fe/H] is rather
difficult to determine. Therefore, we only show the $2\sigma$ and $3\sigma$ 
confidence regions in Fig.~\ref{fig-alpha}. It can be seen that both the solid
curve and the confidence regions are offset from the data on Ti by 
$\sim 0.3$~dex. This may indicate that the Ti yields are systematically
underestimated by CCSN models for some unknown reasons. It is also
possible that the yield ratios of Ti to Fe have much larger variations among
CCSNe than estimated by current models. We have examined the effects
on [Ti/Fe] vs. [Fe/H] by including SN Ia contributions to both Ti and Fe based 
on several models \cite{maeda} and found that the problem persists or even
worsens. Nevertheless, it is worthwhile to explore SN Ia yields of Ti and
Fe with more models. Finally, observational uncertainties should be
examined because there also appears to be a systematic offset between 
Ti abundances determined with Ti I and Ti II in different ionization states.

As can be seen clearly from Fig.~\ref{fig-alpha}, the $2\sigma$ confidence 
regions shrink significantly at [Fe/H]~$\sim-1.4$, especially for Mg, Si, and
Ca. This results from the relatively efficient mixing of gas during the 
pre-satellite phase of Fornax and is consistent with the data. However, 
large scatter in the data appears again at higher [Fe/H] values. 
We take this to indicate that mixing of gas was very inefficient during the 
satellite phase of Fornax, possibly due to the disruption of individual growing 
SN bubbles by the global gas outflow.

This work was supported in part by the US DOE (DE-FG02-87ER40328), 
the NSFC (11320101002), 973-Project 2015CB857000, Shanghai Key
Laboratory Grant No. 11DZ2260700, and the CAS/SAFEA International 
Partnership Program for Creative Research Teams (KJCX2-YW-T23).

\bibliography{omeg15_Yuan}

\begin{thebibliography}{17}

\bibitem{deboer-f}
T.J.L. {de Boer}, E.~{Tolstoy}, V.~{Hill}, A.~{Saha}, E.W. {Olszewski},
  M.~{Mateo}, E.~{Starkenburg}, G.~{Battaglia}, M.G. {Walker}, \aap
  \textbf{544}, A73 (2012)

\bibitem{roychow}
S.~{Roychowdhury}, M.L. {Huang}, G.~{Kauffmann}, J.~{Wang}, J.N. {Chengalur},
  \mnras \textbf{449}, 3700 (2015)

\bibitem{zhao}
D.H. {Zhao}, Y.P. {Jing}, H.J. {Mo}, G.~{B{\"o}rner}, \apj \textbf{707}, 354
  (2009)

\bibitem{wolf}
J.~{Wolf}, G.D. {Martinez}, J.S. {Bullock}, M.~{Kaplinghat}, M.~{Geha}, R.R.
  {Mu{\~n}oz}, J.D. {Simon}, F.F. {Avedo}, \mnras \textbf{406}, 1220 (2010)

\bibitem{yuan}
Z.~{Yuan}, Y.Z. {Qian}, Y.P. {Jing}, ArXiv e-prints  (2015),
  \texttt{1503.00780}

\bibitem{yuan-th}
Z.~{Yuan}, Ph.D. thesis, University of Minnesota (2015)

\bibitem{tol09}
E.~{Tolstoy}, V.~{Hill}, M.~{Tosi}, \araa \textbf{47}, 371 (2009)

\bibitem{kroupa}
P.~{Kroupa}, \mnras \textbf{322}, 231 (2001)

\bibitem{maoz}
D.~{Maoz}, F.~{Mannucci}, T.D. {Brandt}, \mnras \textbf{426}, 3282 (2012)

\bibitem{heger}
A.~{Heger}, S.E. {Woosley}, \apj \textbf{724}, 341 (2010)

\bibitem{maeda}
K.~{Maeda}, F.K. {R{\"o}pke}, M.~{Fink}, W.~{Hillebrandt}, C.~{Travaglio}, F.K.
  {Thielemann}, \apj \textbf{712}, 624 (2010)

\bibitem{thornton}
K.~{Thornton}, M.~{Gaudlitz}, H.T. {Janka}, M.~{Steinmetz}, \apj \textbf{500},
  95 (1998)

\bibitem{kgf15}
A.C. {Petit}, M.R. {Krumholz}, N.J. {Goldbaum}, J.C. {Forbes}, \mnras
  \textbf{449}, 2588 (2015)

\bibitem{letarte}
B.~{Letarte}, V.~{Hill}, E.~{Tolstoy}, P.~{Jablonka}, M.~{Shetrone}, K.A.
  {Venn}, M.~{Spite}, M.J. {Irwin}, G.~{Battaglia}, A.~{Helmi} et~al., \aap
  \textbf{523}, A17 (2010), \texttt{1007.1007}

\bibitem{kirby}
E.N. {Kirby}, G.A. {Lanfranchi}, J.D. {Simon}, J.G. {Cohen}, P.~{Guhathakurta},
  \apj \textbf{727}, 78 (2011)

\bibitem{lemasle}
B.~{Lemasle}, T.J.L. {de Boer}, V.~{Hill}, E.~{Tolstoy}, M.J. {Irwin},
  P.~{Jablonka}, K.~{Venn}, G.~{Battaglia}, E.~{Starkenburg}, M.~{Shetrone}
  et~al., \aap \textbf{572}, A88 (2014)

\bibitem{hendricks}
B.~{Hendricks}, A.~{Koch}, M.~{Walker}, C.I. {Johnson}, J.~{Pe{\~n}arrubia},
  G.~{Gilmore}, \aap \textbf{572}, A82 (2014)

\end{thebibliography}
\end{document}